\documentclass[journal]{IEEEtran}
\usepackage{amssymb}
\usepackage{times,amsmath,epsfig,algorithmic,algorithm,color}

\ifCLASSINFOpdf
\else
\fi
\begin{document}
%
\title{\huge Distributed Iterative Detection Based on Reduced Message Passing for Networked MIMO Cellular Systems}
%
%
%

\author{Peng~Li,~\IEEEmembership{Member,~IEEE}
        and Rodrigo~C.~de~Lamare,~\IEEEmembership{Senior Member,~IEEE} 

\thanks{{Peng~Li was and Rodrigo~C.~de.~Lamare is with the Department of Electronics, The University of York, England UK, YO10 5DD. e-mail: (pl534,rcdl500)@ohm.york.ac.uk. Peng~Li is now with the Communications Research Laboratory, TU-Ilmenau, Germany. Rodrigo~C.~de.~Lamare is also with CETUC,  Pontifical Catholic University of Rio de Janeiro (PUC-Rio), Brazil. }}
}

%
%

\markboth{  {\bf{IEEE Transactions on Vehicular Technology}}, 2014}%
{Shell \MakeLowercase{\textit{et al.}}: Bare Demo of IEEEtran.cls for Journals}

%



\maketitle

\begin{abstract}


This paper considers base station cooperation (BSC) strategies for
the uplink of a multi-user multi-cell high frequency reuse scenario
where distributed iterative detection (DID) schemes with soft/hard
interference cancellation algorithms are studied. The conventional
distributed detection scheme exchanges {soft symbol estimates} with
all cooperating BSs. Since a large amount of information needs to be
shared via the backhaul, the exchange of hard bit information is
preferred, however a performance degradation is experienced. In this
paper, we consider a reduced message passing (RMP) technique in
which each BS generates a detection list with the probabilities for
the desired symbol that are sorted according to the calculated
probability. The network then selects the best {detection
candidates} from the lists and conveys the index of the
constellation symbols (instead of double-precision values) among the
cooperating cells. The proposed DID-RMP achieves an
inter-cell-interference (ICI) suppression with low backhaul traffic
overhead compared with {the conventional soft bit exchange} and
outperforms the previously reported hard/soft information exchange
algorithms.

%

\end{abstract}

\begin{IEEEkeywords}
Distributed iterative detection, multiuser detection, MIMO, base station cooperation, iterative (Turbo) processing.
\end{IEEEkeywords}

%
\IEEEpeerreviewmaketitle

\section{Introduction}
%
%
%
%
The growing demand for mobile multimedia applications requires
higher data rates and reliable links between base stations and
mobile users. The improvement of system capacity can be achieved by
introducing a higher frequency reuse and micro cell planning
\cite{RE4,RE2}. In such a network configuration, a higher spectral
efficiency is obtained, however, the inter-cell interference (ICI)
becomes dominant at the cell edges, especially in an aggressive
frequency reuse scenario \cite{RE4,DMP}. The application of
interference mitigation techniques is necessary in these systems to
prevent a reduced data rate of the users located at the cell edge
and improve the system fairness \cite{LL12,LS08}.

Strategies to deal with the ICI in the system uplink include joint
multiuser detection (JMD) \cite{DMP,CAY05,CA07,LS07} and
{distributed iterative detection (DID)
\cite{KRF08,LL12WSA,KF07,RE1,RE3}}. In terms of JMD, the BSs for
each cell make the received signals available to all cooperating
cells. With this setting, the receivers not only use the desired
signal energy but also the energy from the interferers leading to
much improved received SINR. Both array and diversity gains are
obtained resulting in substantial increase in system capacity
\cite{KF07}. Despite the optimality of JMD, it needs to exchange all
the quantized received signals between the cooperative BSs via a
wired or microwave backbone network which brings about huge
background data traffic \cite{RE2,KRF08}. In order to reduce the
backhaul traffic, clusters may be applied, a group of BSs can form a
cluster and the JMD can be performed in a central unit. The
information is exchanged within the cluster which reduces the
backhaul and the complexity. However the JMD based structure has
many restrictions: (1) The performance degrades at the boundaries of
the clusters. (2) The central units are required to support a large
number of users in the cluster which introduces a high detection
complexity. (3) It requires transmission of quantized received
signals over the wired network to the central unit which causes a
high backhaul traffic \cite{RE4,KRF08}.

In order to circumvent the aforementioned problems, an advanced
interference {mitigation} technique for distributed receivers is
introduced. A DID structure is presented as an alternative to JMD
for cooperative detection with affordable backhaul traffic between
cooperating BSs \cite{MJH06,KRF08,KF07}. With the DID scheme,
iterative processing is performed at the network level. The receiver
detects each user stream in its corresponding cell and iteratively
refines the estimate of the transmitted symbol with the help of the
information provided by other cooperating cells. Each BS detects the
desired user/stream only, the other interfering signal is cancelled
or treated as noise \cite{WP99,HB03}. The output of the receiver is
used to reconstruct the transmitted symbol and this estimate is
conveyed to the cooperating BSs. Each BS exchanges its estimates
with the neighbours, the reconstructed interferers are cancelled
from the received signal and the power of the interference reduces
as more iterations are performed. With DID the detection complexity
is restricted to the number of data streams inside the cell
\cite{KRF08}. Despite their advantages, DID techniques have a
drawback that the interference cancellation is performed at the
network level, the exchange of soft information brings about a high
backhaul traffic and the iterative detection delay must be
minimised.

In the remaining part of this paper, we focus on interference
mitigation techniques \cite{LL12WSA,LL12,LS08} dealing with the
multiuser multicell detection through base station cooperation (BSC)
in an uplink, interference-limited, aggressive frequency reuse
scenario. In the proposed DID with reduced message passing (DID-RMP)
algorithm, the cooperating BSs exchange the information while
performing interference mitigation based on single or multi-user
detection. Instead of exchanging the soft estimates introduced in
\cite{KRF08,WP99,KF07}, the proposed algorithm generates a sorted
list containing the probability of the constellation symbols given
the channel information. The indices of the constellation symbols
with high probability are exchanged via the backhaul link. A
selection unit (SU) is also proposed in the network to provide the
best candidates from the list. The {indices are exchanged} among the
BSs in an iterative manner and the system improves the estimate of
the desired signal with {each iteration loop}. The indexed
interference at the cooperating BSs is subtracted from the received
signal resulting in a reduced interference level and more reliable
data estimates. The simulation results indicate that the proposed
DID-RMP scheme is able to outperform the soft symbol cancellation
technique reported in \cite{KRF08,KF07} while requiring much less
backhaul traffic.

This paper is structured as follows. The system and data model is
presented in section II. In section III, the iterative detection
with RMP is discussed which also involves soft/hard interference
subtraction and the proposed index based subtraction. The simulation
results and the conclusions are presented in sections IV and V
respectively.

\section{Data Model of a Networked MIMO Cellular System}

We consider an asymmetric multiuser scenario of a networked MIMO
cellular system. We assume that the cellular network can detect
groups of users that are received by several cooperating BSs
\cite{MJH06, KRF08, LL12WSA}. We consider that {a number of $\phi$
cells} are grouped into one cluster, the diversity and array gains
can be obtained inside the cluster, the interference among the
clusters is mitigated through the application of DID schemes. Since
we are interested in mitigating the inter-cluster interference, in
order to simplify our description, we consider the special case
$\phi = 1$, where each cell represents a cluster. The scenarios with
more cells in the cluster $\phi > 1$ are straightforward.

Let us consider an idealized synchronous uplink single-carrier
narrowband cellular network that aims to capture most of the
features of a realistic wireless system with respect to the
interference and the need for backhaul. We define $M$ as the number
of {cooperating BSs} and $K$ as the number of users in the
cooperating cells, and assume the users and BSs have a single
transmit antenna. Extensions to multiple antennas are
straightforward and are considered later on. In networked MIMO
systems, a limited number of cells can work together in order for
the backhaul overhead to be affordable \cite{V07}, {by increasing
the number of cooperating cells, a higher number of interfering
links are expected to be dealt with. The increased backhaul traffic
is a direct consequence of the BSs dealing with a higher number of
interferers. Therefore, the number of cooperating cells should be
limited.} In this system, the transmitted data of each user are
protected by the channel codes separately. A message vector
$\boldsymbol{m}_k$ from user $k$ is encoded by a channel code before
a bit interleaving operation. The resulting bit-sequence
$\boldsymbol{b}_k$ has $Q$ entries and $k = 1, 2, \ldots, K$ are
indices of the interfering users. The sequence is then divided into
groups of $J$ bits each, which are mapped to a complex symbol vector
as the output of the user $k$, this operation is denoted as
$\boldsymbol{s}_k = [s_{k,1},\ldots,s_{k,Q_s}] =
\mbox{map}(\boldsymbol{b}_k)$ where $Q_s = Q/J$ and each entry of
$\boldsymbol{s}_k$ is taken from a complex constellation
$\mathcal{A}$ {with power $E\{|s_{k,j}|^2\} = \sigma_s^2$}.

\subsection{Data Model for Single Antenna Users and BSs}

A $K \times 1$ symbol vector $\boldsymbol{s}[i] =
\big[s_{1}[i],s_{2}[i],\ldots,s_{K}[i]\big]^T$ is transmitted
simultaneously by all K users. At base station $m$, the  received
symbols $r_m[i]$ are given by
\begin{equation}\label{transmisson}
  {r}_m[i] = \boldsymbol{g}_m[i]\boldsymbol{s}[i] + v_m[i], \qquad 1\leq i \leq Q_s.
\end{equation}
where $\boldsymbol{g}_m[i] \in \mathbb{C}^{1 \times K}, m = 1,
\ldots, M$, the entry $[i]$ is the time index and $v_m[i]$ denotes
the additive zero mean complex Gaussian noise with variance
$E\{{v[i]}{v[i]}^*\} = \sigma_v^2  $.

The entries of the $1 \times K$ row vector $\boldsymbol{g}_m$ are
the element-wise product of $h_{m,k}$ and $\sqrt{\rho_{m,k}}$, where
$h_{m,k}$ is the complex channel realization from the $k$-th user to
the $m$-th BS with independent and identically distributed (i.i.d)
$\mathcal{CN}(0,1)$. The coefficients $\rho_{m,k}$ reflect the path
loss with respect to BS $m$ and user $k$. Similarly to \cite{KRF08},
we separate ${r}_m[i]$ into four terms expressed by

\begin{equation}\label{transmission2}
\begin{aligned}
  r_m[i] &= g_{m,d}s_d[i] + \sum_{n\in \mathcal{C}_m} {g_{m,n}}s_n[i] + \sum_{n\in \hat{\mathcal{C}}_m}{g_{m,o}}s_o[i] + v[i], \\
     & = \sqrt{\rho_{m,d}}h_{m,d}s_d[i] + \sqrt{\rho_{m,n}}\sum_{n\in \mathcal{C}_m}{h_{m,n}}s_n[i] \\ & \quad + \sqrt{\rho_{m,o}}\sum_{n\in \hat{\mathcal{C}}_m}{h_{m,o}}s_o[i] + v[i].
\end{aligned}
\end{equation}
where the first term denotes the desired signal (indexed by $_d$),
the second and the third terms denote the strong interference and
the weak interference (indexed by $_n$ and $_o$, respectively). The
coefficients $\rho_{n}$ and $\rho_{o}$ characterize the channel
gains with strong and weak interferers, respectively. The set of
indices of all strongly received interference at BS $m$ is denoted
as $\mathcal{C}_m$ and the weakly received interference is denoted
as $\hat{\mathcal{C}}_m$.


{It is shown in \cite{MJH06,KRF08} that the strongest interferers dominate the total ICI}. In this model, we {constrain} the number of strongly {received signals to $m_n \leq 5$}. {For example, a system with $K = M = 4$, and the number of strong interferers $\zeta=2$, the weak interference $\rho_{m,o}$ is equal to zero and the desired user is denoted by $\rho_{m,d} = 1$, then the coupling matrix is formed as}

\begin{equation} \label{couplingmatrix}
\boldsymbol{P} = \left[
\begin{array}{cccc}
 1 &\rho_{m,n} &\rho_{m,n} & 0 \\
 0 &1 &\rho_{m,n} & \rho_{m,n} \\
 \rho_{m,n}&0 &1 & \rho_{m,n} \\
\rho_{m,n} &\rho_{m,n} &0 &1
\end{array}
\right].
\end{equation}

{The coupling matrix $\boldsymbol{P}$ is introduced to describe the configuration of an interference model of a multiuser multicell system. Its diagonal values indicate the power of the link between the BS and the user within the local cell. The off-diagonal values denote the power of interfering links between the BS and the interfering users from other cells. The channel realization of the whole cooperative system $\boldsymbol{G}$ is obtained by the element-wise product of $\boldsymbol{P}$ and $\boldsymbol{H}$ with the elements $h_{m,k}$ following i.i.d. $\mathcal{CN}(0,1)$.} \\

In this configuration, we assume the BSs have the ability to know
which cells are the interfering signals coming from. The BS in the
desired cell then notifies the BS of the interfering cells and
obtains the estimated transmit signal from that cell to perform
interference cancellation. The exchanged interfering information is
transmitted via a wired backhaul which connects all the base
stations in the network.

The signal-to-noise ratio (SNR) is defined as the ratio of the
desired signal power at the receiver side and the noise power, which
is mathematically described as $\mathcal{SNR}_d :=
10\log_{10}\frac{E\big{\{}\|h_{m,d}s_d\|^2\big{\}}}{E\{\sigma_v^2\}}$.
Let us also denote the average signal-to-interference ratio (SIR) of
the desired user $k$ as
\begin{equation}
\begin{aligned}
\mathcal{SIR}_d& := 10\log_{10} & \frac{E\big{\{}\|g_{m,d}s_d\|^2\big{\}} }{\sum_{n\in \mathcal{C}_m} E\big{\{}\|{g_{m,n}}s_n\big{\|}^2\} + \sum_{n\in \hat{\mathcal{C}}_l}E\big{\{}\|{g_{m,o}}s_o\|^2\big{\}}}.
\end{aligned}
\end{equation}

\subsection{Data Model for Multiple Antenna Users and BSs}

In this part, a data model for networked MIMO systems in which the
users and BSs are equipped with multiple antennas is discussed. The
scalar ${r}_m[i]$ and vector $\boldsymbol{g}_m[i]$ in
(\ref{transmisson}) are now described in the vector
${\boldsymbol{r}}_m[i]$ and matrix $\boldsymbol{G}_m[i]$ forms,
respectively, as given by
\begin{equation}
 \boldsymbol{r}_m[i] = \boldsymbol{G}_m \boldsymbol{z}[i] + \boldsymbol{v}_m[i],
\end{equation}
where $\boldsymbol{r}_m \in \mathbb{C}^{N_R \times 1}$ is the
received vector for the $m$-th BS and $\boldsymbol{G}_m  \in
\mathbb{C}^{N_R \times K N_T}$ is the combined channel matrix with
$\boldsymbol{G}_m = \big{[} \boldsymbol{G}_{m,1}, \ldots,
\boldsymbol{G}_{m,k}, \ldots, \boldsymbol{G}_{m,K} \big{]}$ where
$\boldsymbol{G}_{m,k} \in \mathbb{C}^{N_R \times N_T}$ denotes the
channel between user $k$ and BS $m$. Note that each user has $N_T$
transmit antennas and each BS has $N_R$ receive antennas. The
quantity $\boldsymbol{z} \in \mathbb{C}^{KN_T \times 1}$ is the
collection of the data streams from the $K$ users $\boldsymbol{z} =
[\boldsymbol{s}_1^T,\ldots, \boldsymbol{s}_K^T]^T$ and
$\boldsymbol{s}_k \in  \mathbb{C}^{N_T \times 1}$. Equation
(\ref{transmission2}) can be rewritten as
\begin{equation}
\begin{aligned}
  \boldsymbol{r}_m[i] & = \boldsymbol{G}_{m,d}\boldsymbol{s}_d[i] + \sum_{n\in \mathcal{C}_m} {\boldsymbol{G}_{m,n}}\boldsymbol{s}_n[i] + \sum_{n\in \hat{\mathcal{C}}_m}{\boldsymbol{G}_{m,o}}s_o[i] + \boldsymbol{v}_m[i], \\
     & = \sqrt{\rho_{m,d}}\boldsymbol{H}_{m,d}\boldsymbol{s}_d[i] + \sqrt{\rho_{m,n}}\sum_{n\in \mathcal{C}_m}{\boldsymbol{H}_{m,n}}\boldsymbol{s}_n[i] \\
     & \quad + \sqrt{\rho_{m,o}}\sum_{n\in \hat{\mathcal{C}}_m}{\boldsymbol{H}_{m,o}}\boldsymbol{s}_o[i] + \boldsymbol{v}_m[i],
\end{aligned}
\end{equation}
where we assume that the $N_T$ antennas for each user have the same
channel gain coefficients $\rho_n$ and $\rho_o$. The coupling matrix
given in (\ref{couplingmatrix}) and the definition of SNR and SIR
can be generalized accordingly.

In order to simplify the description of the proposed structure and
its traditional counterparts, we first employ the single antenna
case $N_T = N_R = 1$ in the following section.

\section{Distributed Iterative Detection with Reduced Message Passing}

In this section, the decision-aided DID structure is described in
detail, in the first subsection the distributed iterative signal
processing in an interference limited cellular network is reviewed.
In the following two subsections, the soft and hard parallel
interference cancellation algorithms are based on the quantized
estimates from the cooperating BSs. The last subsection is devoted
to the description of the proposed DID-RMP.

\subsection{Decision-Aided Distributed Iterative Detection}
\begin{figure}[h]{
\centering \mbox{\epsfxsize=3.5in \epsffile{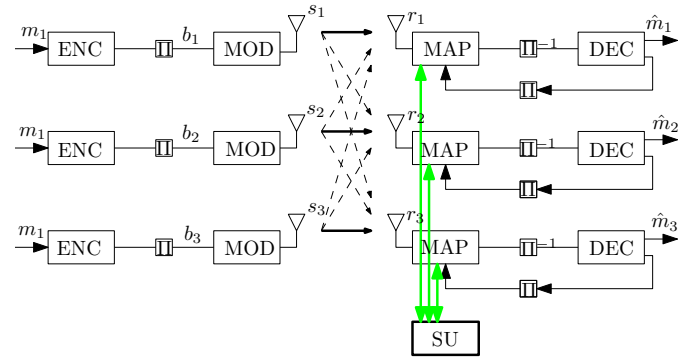}} \caption{An
example configuration showing a cooperating 3-cell network. The
dashed lines between the transmitter and receiver denote the ICI
while the solid lines denote the desired signal.}\label{fig:5_IDD}}
\end{figure}

The setup for performing the distributed detection with the
information exchange between base stations is shown in
Fig.{\ref{fig:5_IDD}}. The $K$ users' data are separately coded and
modulated to complex symbols after bit-interleaving. At each BS, the
received signal $r_m[i]$ is the collection of the transmitted signal
and the Gaussian noise.

In addition, each BS equips a communication interface for exchanging
information with the cooperating BSs. The information is in the form
of a bit sequence that represents the quantized soft estimates. The
interface is capable of transmitting and receiving information. Via
these interfaces, each cooperating BS is connected to a device,
namely the selection unit (SU), and is ready to receive and transmit
the information for cooperation. The proposed SU has very limited
computational power and it can be integrated with BSs in the
network.

In each BS, a block of received signals $r_m[i]$ is used by the MAP
demapper to compute the \textit{a posteriori} probability in the
form of log-likelihood-ratios (LLRs), which are given by
\begin{equation}
 \Lambda_1^p[b_{j,k}[i]] = \log {\frac{P[b_{j,k}[i] = +1 | r_m[i] ]}{P[b_{j,k}[i] = -1 | r_m[i]]}},
\end{equation}
where the equation can be solved by using Bayes' theorem and we
leave the details to the references \cite{WP99,HB03}. The detector
and the decoder are serially concatenated to form a ``turbo"
structure, the \textit{extrinsic} information is exchanged by the
two soft-input soft-output components. We denote the
\textit{intrinsic} information provided by the decoder as
$\Lambda_2^p[b_{j,k}[i]]$ and the bit probability is $P[b_{j,k}[i]]
= \log\frac{P[b_{j,k}[i] = +1]}{P[b_{j,k}[i] = -1]}.$ From
\cite{WP99}, the bit-wise probability is obtained by
\begin{equation}\label{bitp}
\begin{aligned}
P[b_{j,k}[i] = \bar{b}_j] & =
\frac{\exp\Big{(}{\bar{b}_j}\Lambda_2^p[b_{j,k}[i]]\Big{)}}{1+\exp\Big{(}{\bar{b}_j}\Lambda_2^p[b_{j,k}[i]]\Big{)}}
\\ & =
\frac{1}{2}\left[1+\bar{b}_j\tanh\Big{(}\frac{1}{2}\Lambda_2^p[b_{j,k}[i]\Big{)}\right],
\end{aligned}
\end{equation}
where $\bar{b}_j = \{ +1, -1\}$. Let us simplify the notation
$P\big[s_k[i]\big] := P\big[s_k[i] = {c}_q\big]$ where $c_q$ is an
element chosen from the constellation $\mathcal{A} =
\{c_1,\ldots,c_q,\ldots,c_A\}$. The symbol probability $P[s_k[i]]$
is obtained from the corresponding bit-wise probability, and
assuming the bits are statistically independent, we have
\begin{equation}\label{sybp}
\begin{aligned}
 P\big[s_k[i]\big] & = \prod_{j = 1}^{J} P\big[b_{j,k}[i] = \bar{b}_j\big] \\ & = \frac{1}{2^J} \prod_{j = 1}^{J} \left[1+\bar{b}_j\tanh\Big{(}\frac{1}{2}\Lambda_2^p[b_{j,k}[i]\big{]}\Big{)}\right].
\end{aligned}
\end{equation}
From (\ref{bitp}) and (\ref{sybp}) we can easily conclude that
$\sum_{|\mathcal{A}|} P\big[s_k[i]\big] = 1.$ The symbol likelihood
$ P\big[s_k[i]\big]$ can be used to evaluate the reliability of the
recovered symbol. A higher probability of detection of $s_k[i]$ can
be associated with a higher reliability of estimation of that
symbol.

\subsection{Soft Interference Cancellation}

The soft interference cancellation has first been reported in an
iterative multiuser CDMA systems by Wang \textit{et al} in
\cite{WP99} and later extended by several works
\cite{MJH06,HB03,RE3}. In the algorithm \cite{MJH06}, the soft
replicas of ICI are constructed and subtracted from the received
signal vector as
\begin{equation}\label{eq:sc}
 \tilde{{r}}_{m,k}[i] = {r}_{m}[i] - \boldsymbol{g}_m
 \tilde{\boldsymbol{u}}_k[i]
\end{equation}
and the replica of the transmitted symbol vector $
\tilde{\boldsymbol{u}}_k[i] \in \mathbb{C}^{K \times 1}$ is obtained
as
\begin{equation}\label{eq:replica}
 \tilde{\boldsymbol{u}}_k[i] = \Big[\tilde{s}_1[i], \ldots, \tilde{s}_{k-1}[i], 0, \tilde{s}_{k+1}[i], \ldots, \tilde{s}_{K}[i] \Big{]}^T,
\end{equation}
where the estimates of $s_k[i]$ are calculated as
\begin{equation}
 \tilde{s}_k[i] = E\{s_k[i]\} = \sum_{c_q \in \mathcal{A}} c_q P\big[s_k[i] = c_q\big].
\end{equation}
The first-order and second-order statistics of the symbols are
obtained from the symbol a priori probabilities as
$\sigma_{\scriptsize \mbox{eff}}^2 = \mbox{var}\{s_k[i]\} =
E\Big{\{}\big|s_k[i]\big|^2\Big{\}} - \big|\tilde{s}_k[i]\big|^2$
and $ E\Big{\{}\big|s_k[i]\big|^2\Big{\}} = \sum_{c_q \in
\mathcal{A}} |c_q|^2 P\big[s_k[i] = c_q\big]$.

In the case that the users and BSs are equipped with multiple
antennas, then (\ref{eq:sc}) can be reformulated as
\begin{equation}
 \tilde{\boldsymbol{r}}_{m,k}[i] = \boldsymbol{r}_{m}[i] - \boldsymbol{G}_m \tilde{\boldsymbol{u}}_k[i].
\end{equation}

{The soft interference cancellation procedure can be considered in
two cases. In the first case, the cancellation is performed in terms
of users rather than data streams, and we name this case as
user-based cancellation. In this case, the interfering signals
received from other cell users are canceled but the interference
between the antenna data streams of the desired user remains.
Mathematically, the replica of the transmitted symbol vector}
$\tilde{\boldsymbol{u}}_k[i] \in \mathbb{C}^{KN_T \times 1}$  is
defined as
\begin{equation}
\tilde{\boldsymbol{u}}_k[i] = \Big[\tilde{\boldsymbol{s}}_1^T[i],
\ldots, \tilde{\boldsymbol{s}}_{k-1}^T[i], \boldsymbol{0},
\tilde{\boldsymbol{s}}_{k+1}^T[i], \ldots,
\tilde{\boldsymbol{s}}_{K}^T[i] \Big{]}^T,
\end{equation}
where $\boldsymbol{0} \in \mathbb{Z}^{N_T \times 1}$ and
$\tilde{\boldsymbol{s}}_{\kappa \neq k}^T[i], \kappa = 1, \ldots, K.
\in \mathbb{C}^{N_T \times 1} $. The remaining signal after the
interference cancellation is the combination of all the data streams
transmitted from user $k$ and the noise.

{In the second case, we consider each independent antenna data
stream received by the BSs and disregarding which users send them,
we name this case as data stream-based cancellation. In this case,
the BSs consider interference in terms of streams instead of users.
In a mathematical point of view, the replica of the transmitted
signal for stream-based interference cancellation
$\tilde{\boldsymbol{u}}_k[i] \in \mathbb{C}^{KN_T \times 1}$ is
defined as
\begin{equation}
\tilde{\boldsymbol{u}}_k[i] = \Big[\tilde{\boldsymbol{s}}_1^T[i], \ldots, \tilde{\boldsymbol{s}}_{k-1}^T[i],  \tilde{\boldsymbol{s'}}_{k}^T[i], \tilde{\boldsymbol{s}}_{k+1}^T[i], \ldots, \tilde{\boldsymbol{s}}_{K}^T[i] \Big{]},
\end{equation}
where the entry $\tilde{\boldsymbol{s'}}_{k}$ is obtained as
$\tilde{\boldsymbol{s'}}_{k}^T[i] = \Big[\tilde{s}_1[i], \ldots,
\tilde{s}_{n_t-1}[i], 0, \tilde{s}_{n_t+1}[i], \ldots,
\tilde{s}_{N_T}[i] \Big{]}^T$. By using this scheme, all the
interfering streams are removed after the cancellation procedure.}

This soft interference cancellation based algorithm generally
outperforms hard interference cancellation since it considers the
reliability of the cancellation procedure. However, the performance
heavily depends on the quantization level. Exchanging the quantized
soft bits or LLRs convey reliability information among BSs and
involves a large amount of backhaul data per cell per iteration,
which make soft interference cancellation unattractive.

\subsection{Hard Interference Cancellation}
With the hard interference cancellation, the estimates of the
interfering symbols are the constellation symbols. In this case, the
quantization is performed for each estimated symbol. Equation
(\ref{eq:replica}) is rewritten as
\begin{equation}
 \hat{\boldsymbol{u}}_k[i] = \Big[\mbox{Q}(\tilde{s}_1[i]), \ldots, \mbox{Q}(\tilde{s}_{k-1}[i]), 0,\mbox{Q}({\tilde{s}_{k+1}}[i]), \ldots, \mbox{Q}({\tilde{s}_{K}}[i]) \Big{]}^T,
\end{equation}
where $\mbox{Q}(\cdot)$ is the slicing function that depends on the
constellation adopted. The constellation indices are exchanged among
the cooperating BSs. Since no reliability information is included,
the cooperation procedure requires significantly less backhaul
traffic as compared with the soft interference procedure. All the
detected information symbols are exchanged in the initial iteration,
and in the subsequent iterations, only the symbols with the
constituent bits that have flipped between the iterations are
exchanged.  The indexed constellation symbols are reconstructed at
the neighboring BSs and subtracted from the received signal, the
residual noise is considered equal to zero and $\sigma_{\scriptsize
\mbox{eff}}^2 = \sigma_v^2$. In the hard interference cancellation
configuration, the backhaul traffic can be further brought down by
introducing a reliability check of the symbols and by exchanging
reliable symbols. {It is worth to mention that by introducing the
reliability check, the error propagation effect can be effectively
mitigated. The selected unreliable estimates can be either refined
or excluded from the interference cancellation procedure. The
performance improvement over the hard IC scheme is investigated in
\cite{LLF11}}.

\subsection{Distributed Iterative Detection with Reduced Message Passing }
The hard interference cancellation is performed in a way that the
effect of all the detected symbols but the intended one are removed
from the received signal. It ignores the reliability of the
estimated symbols used for interference cancellation, but ignoring
the reliability may lead to error propagation, which can
significantly deteriorate the performance. The soft interference
cancellation is then introduced to combat error propagation by using
quantized soft symbols, however, this procedure requires more
iterations to obtain a good performance {which increases the
detection delay.} In additional, the sharing of quantized symbol
estimates requires a higher bandwidth across the network and the
bit-wise quantization for every symbol brings about a higher
complexity. In the following subsection, we present a method which
is able to {address these problems} and keep a low backhaul
requirement.

By organizing the probabilities obtained by (\ref{sybp}) in
decreasing order of values, a list of tentative decisions of
$s_k[i]$ is obtained in each BS, as given by
\begin{equation}\label{rmp1}
 \mathcal{L}_k[i] \triangleq \{c_1, c_2, \ldots, c_\tau\}_k,
\end{equation}
{where the number of candidates is $1\leq \tau \leq |\mathcal{A}|$}.
The probabilities $Pr[c_1] \geq Pr[c_2] \geq \ldots, Pr[c_\tau]$
where $Pr{[} c_q] \triangleq P\big{[}s_k[i] =
c_q\big{|}{r_m}\big{]}$ is the probability of the transmitted signal
is $c_q$ given $r_m$. {For the sake of simplicity of computation, we
only keep candidates with probability higher than a threshold such
as $P\big[s_k[i]\big] \geq \rho_{\scriptsize \mbox{th}}$ from the
list. The threshold $\rho_{\scriptsize \mbox{th}}$ may be fixed or
varied in terms of SINR.} {It is also worth to mention that failing
to optimize the threshold $\rho_{\scriptsize \mbox{th}}$ would
result in either heavy backhaul traffic ($\rho_{\scriptsize
\mbox{th}}$ too low) or unacceptable performance ($\rho_{\scriptsize
\mbox{th}}$ too high). The optimization of $\rho_{\scriptsize
\mbox{th}}$ can be performed by maximizing the SINR of the data
streams with the constraint of the maximum allowable backhaul
traffic.}

{For symbols transmitted by each user}, we generate a tentative
decision list $\mathcal{L}_k$. By listing all the combinations of
the elements across $K$ users, a length $\varGamma$ tentative
decision list is formed at the corresponding SU. {Each column vector
on the list denotes a possible symbol vector $\boldsymbol{s}'_l$
where $l = 1, \ldots, \varGamma$.} The size of the list is obtained
by
\begin{equation}
 \varGamma = \prod_{k=1}^{K}|\mathcal{L}_k|,  \qquad 1\leq\varGamma\ll |\mathcal{A}|^K,
\end{equation}
where $|\cdot|$ denotes cardinality. In order to obtain an improved
performance, the maximum likelihood (ML) rule can be used to select
the best among the $\varGamma$ candidate symbol vectors. {Note that
without a designated threshold, an ML search over the whole vector
space $\Gamma = |\mathcal{A}|^K$ is performed, which is equivalent
to joint ML detection and provides a full diversity order with
prohibitive backhaul requirements and detection complexity. However,
the DID-RMP algorithm obtains a higher diversity order than that of
``perfect interference cancellation" with a much smaller candidate
list (compared with ML) thanks to the threshold $\rho_{\scriptsize
\mbox{th}}$ and its effective selection of candidates.

The threshold value should be adequately set in order to generate an
affordable list size $\Gamma$.} The ML criterion, which is
equivalent to the minimum Euclidean distance criterion, computes the
ML solution as given by
\begin{equation}\label{eq:eucli}
 \boldsymbol{s}'_{\scriptsize \mbox{ML}} = \arg\min_{l = 1,\ldots,\varGamma} \big{\|}\boldsymbol{r}[i] - \boldsymbol{G}\boldsymbol{s}_l'[i]\big{\|}^2,
\end{equation}
where $\boldsymbol{r}[i] = [r_1[i], \ldots, r_m[i], \ldots, r_M[i]]^T$ and { $\boldsymbol{G} = [\boldsymbol{g}_1^T, \ldots, \boldsymbol{g}_m^T, \ldots, \boldsymbol{g}_M^T]^T$} are  received signals and the user channels for all cooperating cells.

In the above expression, the knowledge of {$\boldsymbol{g}_m$} and
the received signal $r_m[i]$ for each cell is required to be passed
to the SU which may lead to high backhaul traffic. Additionally, as
a central point, there is high computational power demand for the SU
{to choose} the best candidate from the list. In order to circumvent
the aforementioned problems, we introduce the method of reduced
message passing which is able to distribute the normalization
operations {to each cooperating BSs}.

\subsubsection*{Distributed Selection Algorithm}
The Euclidean distance $\boldsymbol{d} = \boldsymbol{r}[i] - \boldsymbol{G}\boldsymbol{s}_l'[i]$ in (\ref{eq:eucli}) is obtained by
\begin{equation}
 \|\boldsymbol{d}\| \triangleq \sqrt{|d_{1,m}|^2+\cdots+|d_{K,m}|^2},
\end{equation}
where $d_{k,m} =  r_m[i] - {\boldsymbol{g}_m}  \boldsymbol{s}'_l[i]$, $\boldsymbol{g}_m[i] \in \mathbb{C}^{1 \times K}, m = 1, \ldots, M$ and $\boldsymbol{s}'_l[i] \in \mathbb{C}^{K \times 1}$. For each BS, we separately calculate the minimum partial weights by
\begin{equation}\label{eq:mpw}
 l^{\scriptsize \mbox{min}}_m =  \arg \min_l | r_m[i] - {\boldsymbol{g}_m}  \boldsymbol{s}'_l[i] |^2.
\end{equation}
{The channel information $\boldsymbol{g}_m$ is known to the local BS $m$ and the candidate with the minimum Euclidean distance index} $l^{\scriptsize \mbox{min}}_m$ is obtained by the SU via backhaul and an enhanced detection is obtained. In each iteration,  the received signal is subtracted by
\begin{equation}\label{eq:scrmp}
 \tilde{{r}}_k[i] = {r}_k[i] - \boldsymbol{h}_k \tilde{\boldsymbol{u}}_k^{\tiny \mbox{ML}}[i],
\end{equation}
where the {selected candidate $\tilde{\boldsymbol{u}}_k^{\tiny \mbox{ML}}$ consists of}
\begin{equation}
\tilde{\boldsymbol{u}}^{\tiny \mbox{ML}} = \big[\tilde{s}_1^{\tiny \mbox{ML}}, \ldots, \tilde{s}_{k-1}^{\tiny \mbox{ML}}, 0, \tilde{s}_{k+1}^{\tiny \mbox{ML}}, \ldots, \tilde{s}_{K}^{\tiny \mbox{ML}} ].
\end{equation}
With this multiple candidate structure, an enhanced ICI suppression
is obtained. The indices of the symbols on the tentative decision
list $\mathcal{A}_k$ are propagated among the neighboring BSs which
require a reduced backhaul traffic compared with that of the soft
signal cancellation algorithm. Additionally, as more cancellation
iterations are performed, the size of the list reduces as the
recovered bits are more reliable. This further decreases the
backhaul traffic with the following iterations, which is not the
case with the approach that adopts a soft interference cancellation
strategy. {We can translate the proposed DID-RMP algorithm as
follows. In a cooperative network serving several users, if one
estimate is not reliable enough to perform interference
cancellation, the system uses the side information (symbol indices)
provided by other cooperative cells to refine this estimate and
therefore, a more reliable interference cancellation in the network
level is obtained.} The algorithm of the proposed DID-RMP method is
summarized in Table. \ref{alg:DID-RMP}.

For an interference cancellation based method, the performance is
bounded by the BER of isolated cells, the single BS in each cell can
only provide a diversity order of one. On the other hand, in an
extreme case, if the algorithm searches the whole vector space $
\varGamma = |\mathcal{A}|^K$, a full diversity order is obtained and
the optimal detection requires exponentially increased complexity.
The DID-RMP algorithm however provides a tradeoff between
complexity/backhaul and performance by varying the threshold
$\rho_{\scriptsize \mbox{th}}$, and a higher diversity order is
obtained with a short candidate list thanks to its effective
selection of candidates.

\section{Complexity and Backhaul Analysis}

In this section, we detail the computational complexity and the
requirement for backhaul of the proposed DID-RMP technique.

\subsection{Complexity}

In terms of the complexity, {a network wide parallel interference
cancellation} is adopted to remove the co-channel interference by
removing the estimates of the interfering symbols based on the a
priori LLRs obtained from the SISO channel decoder. For each
interference cancellation iteration, the reconstruction operations
(\ref{bitp}) and (\ref{sybp}) require $\mathcal{O}(2J)$ real value
multiplications. These symbol estimates are used to cancel
interference in the receiver vector/scalar (\ref{eq:scrmp}) which
require $\mathcal{O}(K-1)$ complex multiplications. The remaining
term is then detected by a soft output MAP detector, the computation
of per-stream a posteriori LLRs requires $\mathcal{O}(J)$ real value
multiplication and $\mathcal{O}(3JK)$ complex multiplications where
$J$ is the modulation level which denotes the number constituent
bits per symbol and $K$ is the total number of users for detection.

Unlike a centralized methods which requires $\mathcal{O}(J^K)$
complex multiplications or $\mathcal{O}(K^2(MK))$ operations for the
filter based signal processing \cite{LL12,LS08,LS07}, in the
proposed DID-RMP structure, each BS separately calculates the
minimum partial weights in each cell (\ref{eq:mpw}) at the cost of
only $\mathcal{O}(\varGamma K)$ complex multiplications and send the
constellation indices to the SU. Therefore, the SU is used as memory
storage of constellation indices with no computational requirement.
The proposed SU is incorporated to minimize the computational
requirement for the SU and maximize the overall performance across
the cells.

{In order to reduce the detection complexity of the proposed DID-RMP
algorithm, list sphere decoders \cite{HB03} and their variants can
be used to generate this candidate list with much lower complexity
as compared to the optimal ML detector. {Furthermore, the MMSE/ZF
based non-linear detectors can be used to perform iterative
detection as well. The detector first separates the spatially
multiplexed data streams and converts the MMSE estimates into bit
level LLRs, then the procedure of (\ref{rmp1}) - (\ref{eq:eucli})
can be applied. However, for MMSE/ZF based methods, by fixing an
allowable backhaul traffic, a worse BER performance is expected due
to its suboptimal performance}. To address this, the authors suggest
an upgraded version of the {successive interference cancellation
algorithm} called MF-SIC \cite{LLF11} to detect the symbols. This
algorithm considers the reliability of estimated symbols and refine
those unreliable ones. Since this algorithm has a near ML
performance with low complexity, we expect a similar performance
with the ML based decoder introduced here.}

\subsection{Backhaul Requirement}


The backhaul requirement for a conventional cooperating cellular
system with soft information exchange depends on the resolution of
quantization for channel state information, the resolution of
quantization for the signal received from each antennas, the number
of cooperating BSs and the number of strong interferers at receiver
side. Whenever a hard information exchange is adopted, the backhaul
requirement is significantly reduced with {the sacrifice of the
detection performance}. By calculating the minimum partial weights
and exchanging the indices of candidate symbols, DID-RMP introduces
a tradeoff between backhaul requirement and performance.

\begin{figure}{
\centering \mbox{\epsfxsize=3.5in \epsffile{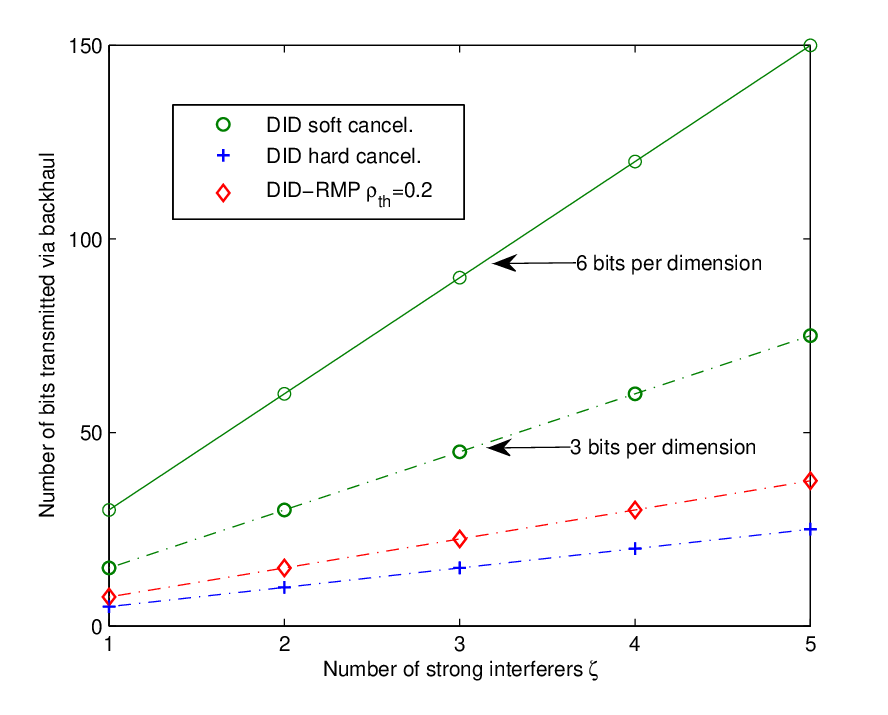}} \caption{
Number of bits exchanged per symbol detection in a 9-cell network.
{The number of bits required via backhaul increases with the number
of strong interfering links within the cooperative
network.}}\label{Fig:backhaul}}
\end{figure}

Fig. \ref{Fig:backhaul} illustrates the backhaul traffic as a
function of the number of strong interferers $\zeta$. As QPSK
modulation is used, 2 bits are required to index the constellation
symbols to perform hard interference cancellation. {In practical
joint and distributed cooperative networks, the data compression
techniques are useful for transmitting the soft quantized symbols.
For the sake of fairness, we compare both 3 bits and 6 bits per
dimension for quantizing the soft symbol, the data compression is
only considered in this section but not in the BER simulations in
the next section.} {With the DID-RMP algorithm, the list size
$\varGamma$ does not grow exponentially with the increase of the
modulation level (e.g. from QPSK to 16QAM), but a higher backhaul
requirement is expected due to an increasing number of unreliable
estimates. On the other hand, if the backhaul reaches its maximum
allowable traffic, performance degradation is also expected.} The
plots indicate that increasing the number of strong interferers for
each cell leads to the rise of the backhaul traffic. Compared with
soft interference cancellation with quantization of the reliability
information algorithm reported in \cite{KRF08}, the proposed DID-RMP
algorithm significantly reduces the backhaul requirement with the
increased number of interferences.

\section{Simulations}
In the simulations, we assume $\rho_{m,o}$ is zero, $\rho_{m,d} = 1$
and  strongly received interference have $\rho_{m,n} = 0.5$. All BSs
are assumed to have the same signal-to-noise ratio (SNR) and the
interfering BSs are also assumed to have the same
signal-to-interference ratio (SIR). In order to evaluate the
performance of the distributed turbo system, we select a rate R =
1/2 convolutional code with polynomial $[7,5]_{\scriptsize
\mbox{oct}}$. The coded bits are modulated as QPSK symbols before
transmission. The decoding is performed by a max-log-MAP decoder and
the block length is set to 1024. The number of detector and decoder
iterations is fixed to 10. The loop of network level interference
cancellation performed by the network stops with iteration 4 and the
number of cells in each cluster is $\phi = 1$, if not otherwise
stated. For the soft interference cancellation scheme \cite{MJH06,
KRF08}, a uniform quantizer is applied in order to quantize the soft
estimates. Without significant information loss compared with the
unlimited backhaul performance, 6 quantization bits per real
dimension backhaul traffic is assumed \cite{KF07}.

\begin{figure}{
\centering \mbox{\epsfxsize=3.5in \epsffile{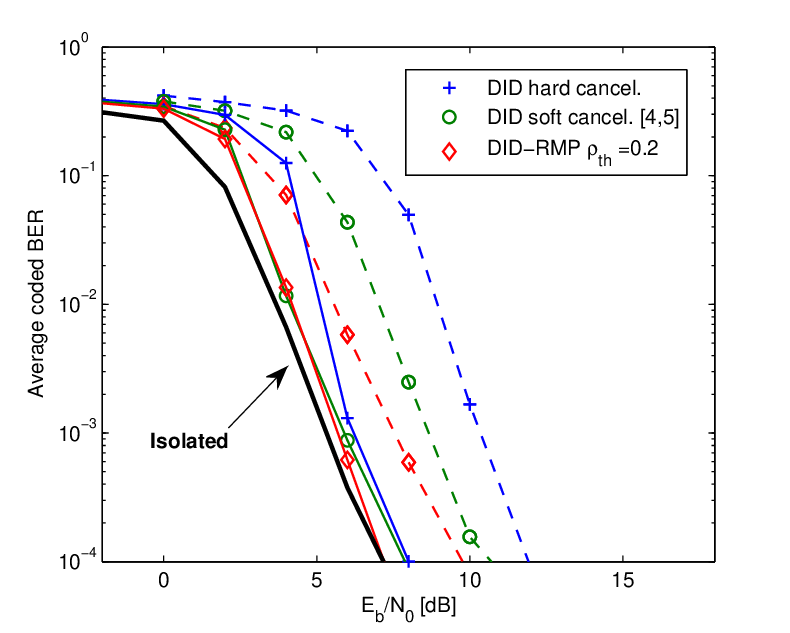}} \caption{SNR
vs. Average BER. The solid lines denote a cooperating 4-cell network
with $\zeta=2$ strong interferers per cell. The dashed lines denote
a cooperating network with 9 cells with $\zeta=3$ strong interferers
per cell. The DID soft cancellation is performed according to
\cite{MJH06,KRF08} }\label{Fig:DIDBER}}
\end{figure}

In Fig. \ref{Fig:DIDBER} the proposed DID-RMP outperforms the soft
interference cancellation  scheme \cite{MJH06,KRF08}, and the
improvement increases with a higher number of strong interferers
$\zeta$. With $\zeta=3$, the proposed scheme achieves about 3 dB of
gain as compared with the system using hard cancellation at the
target $\mbox{BER}=10^{-3}$. {There are 3 dominant interferer at the
BS's receiver. Some weaker interferences below a certain threshold
can be modeled as Gaussian noise and integrated into the noise term.
Therefore, we treat weak interference as noise and the system
considers only strong interference and noise.}

\begin{figure}{
\centering \mbox{\epsfxsize=3.5in \epsffile{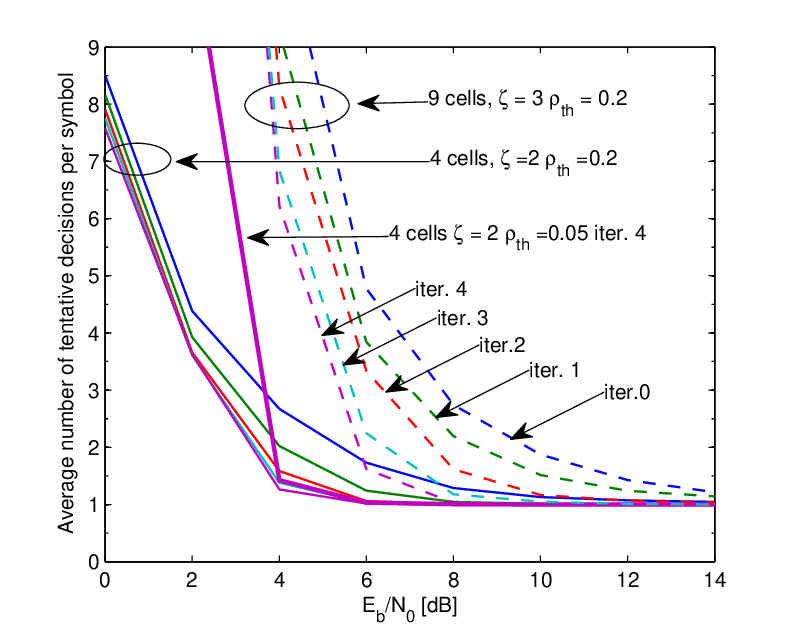}} \caption{The
number of tentative decisions $\varGamma$ decreases as the increase
of SNR. With a smaller threshold $\rho_{\scriptsize \mbox{th}}$
selected, more decision candidates are generated, especially in the
low SNR region.}\label{NCD}}
\end{figure}

In Fig. \ref{NCD}, the average number of tentative decision in the
network is depicted. The number of tentative decisions $\varGamma$
decreases as more iterations are performed. In the proposed DID-RMP
scheme, only indices are exchanged, the backhaul traffic becomes
lower in each iteration due to the fact that $\varGamma$ is getting
smaller. On the other hand, the soft interference cancellation
scheme \cite{MJH06,KRF08} does not benefit from the iterations due
to the requirement of updating the soft estimates. We can also see
from the plots that the average number of candidates quickly
converges to 1, which means low additional detection complexity is
required for each BS. Compared with Fig. \ref{Fig:DIDBER}, the
target BER region from $10^{-3}$ to $10^{-4}$ and the corresponding
SNR is ranged 8 to 10 dB. The average number of tentative decisions
per symbol is below 3 for $\zeta = 3$. In case of two strong
interferers, we can see that negligible additional backhaul overhead
is required.

\begin{figure}{
\centering \mbox{\epsfxsize=3.5in \epsffile{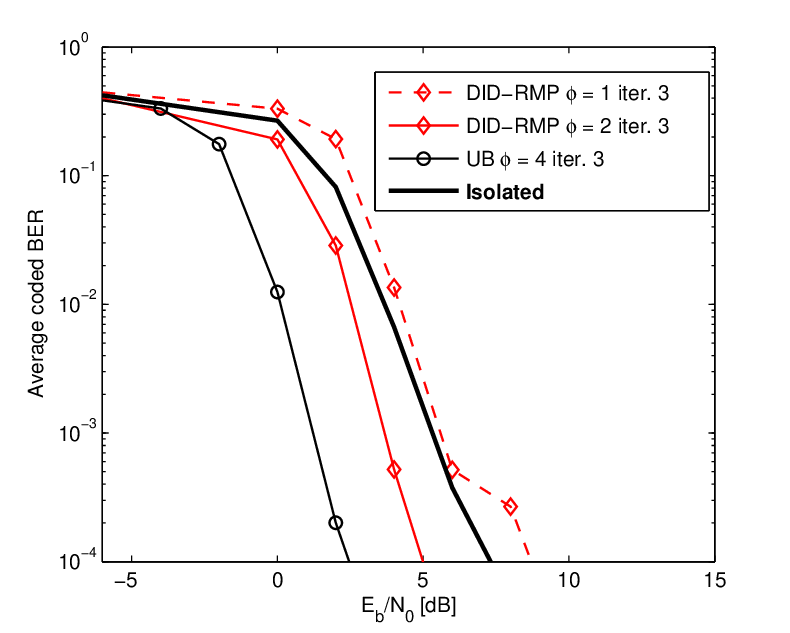}}
\caption{Performance of a cooperating 4-cell network with $\zeta=2$
strong interferers per BS, we group the four cells into two clusters
$\phi = 2$ and single cluster $\phi = 4$.}\label{Fig:BERDiv}}
\end{figure}

All the previous results are bounded by the isolated cell
performance, since $\phi = 1$ and there is only one pair of receive
and transmit antennas available in each cluster, no array gain and
diversity can be obtained. However, in Fig. {\ref{Fig:BERDiv}} we
assume a cooperating 4-cell network with $\zeta=2$ strong
interferers per BS, we group the four cells into two clusters and
$\phi = 2$. A $2 \times 2$ distributed MIMO system is created in
each cluster and the interference is mitigated between two clusters.
We also investigate a single cluster system with $\phi = 4$,
assuming unlimited backhaul (UB), a $4 \times 4$ distributed MIMO
system is created and high diversity and array gain are obtained.

\begin{figure}{
\centering \mbox{\epsfxsize=3.5in \epsffile{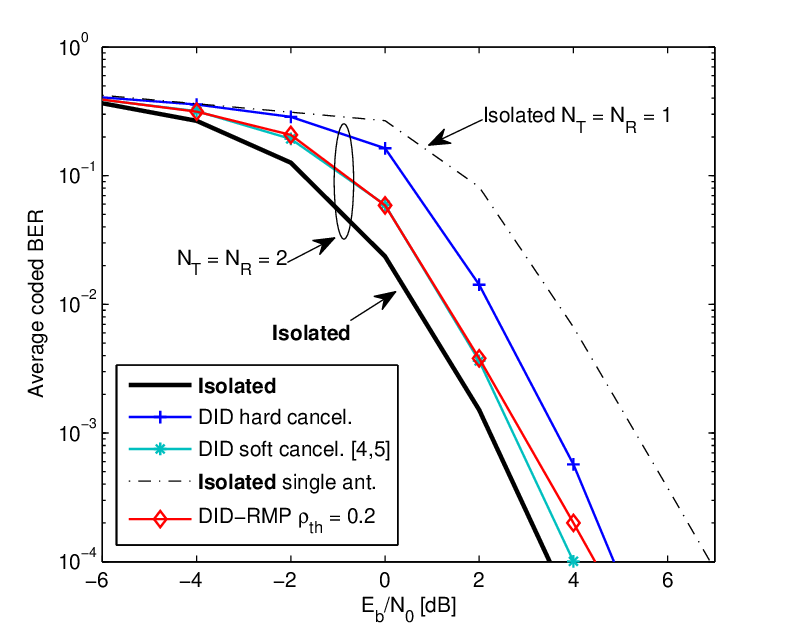}}
\caption{Performance of a cooperating 2-cell network with
$\zeta=\{1,1\}$ strong interferers per BS in which we assume a
single cell for each cluster $\phi = 1$ and $N_R = N_T = 2$ antennas
for each BS and user. {A user-based cancellation is used. The DID
soft cancellation is performed according to
\cite{MJH06,KRF08}.}}\label{Fig:ntnr2}}
\end{figure}

Fig. \ref{Fig:ntnr2} illustrates a system model with
multiple-antenna users and BSs, we build a two cell network model
where each cell has a single user which has $N_T = 2$ transmit
antennas. The BSs for the cells also have $N_R = 2$ antennas ready
for detection. Each BS receives the desired signal as well as the
interference from {the adjacent cells}. Due to the fact that two
data streams are seen as an interfering signal, we use $\zeta =
\{1,1\}$ to discriminate from the single antenna case. {In this
simulation a user-based cancellation is used, the interference
cancellation is only achieved between the users instead of data
streams, the co-channel interference from a single user remains. By
using a fixed threshold $\rho_{\mbox{th}} = 0.2$ for a cooperative
2-cell network with multiple data streams for each user, the DID-RMP
algorithm can provide a near soft-interference cancellation
performance}. 


\vspace{-0.4cm}
\section{Conclusion}

We have discussed multiuser multicell detection through base station
cooperation in an uplink, high frequency reuse scenario. Distributed
iterative detection has been introduced as an interference
mitigation technique for networked MIMO systems.  {We have compared
soft and hard information exchange and cancellation schemes and
proposed a novel hard information exchange strategy based on the
concept of reduced message passing.} The proposed DID-RMP algorithm
significantly reduces the backhaul data compared with the soft
information exchange while it obtains a better BER performance.


%





\ifCLASSOPTIONcaptionsoff
  \newpage
\fi


\begin{algorithm}
\caption{DID-RMP Algorithm}
\label{alg:DID-RMP}
\algsetup{
linenosize=\small,
linenodelimiter=.
}
\begin{algorithmic}[1]
\STATE{\bf Initialization} {$ {r}_m$, $\boldsymbol{g}_m $, $\Lambda_2^p[b_{j,k}[i]] \leftarrow \boldsymbol{0} $, $TI$. }

\FOR{$k \leftarrow 1,\ldots, K$ \COMMENT{\textit{user k}}}
 \STATE $m \leftarrow k$
 \FOR{$j \leftarrow 1,\ldots,J$ \COMMENT{\textit{bit-mapping}}}
    \STATE $P[b_{j,k}[i] = \bar{b}_j] \leftarrow \frac{1}{2}\left[1+\bar{b}_j\tanh\Big{(}\frac{1}{2}\Lambda_2^p[b_{j,k}[i]]\Big{)}\right]$
 \ENDFOR
 \STATE $P\big[s_k[i]\big] \leftarrow \prod_{j = 1}^{J} P[b_{j,k}[i] = \bar{b}_j]$
 \STATE $\mathcal{L}_k[i] \triangleq \{c_1, c_2, \ldots, c_\tau\}_k$ \COMMENT{\textit{candidate list}}
 \STATE {\bf SU $ \boldsymbol{\Leftarrow}  1,\ldots,\tau$}  \COMMENT{\textit{index sharing}}
 \STATE { $ \boldsymbol{s}'_l[i] \boldsymbol{\Leftarrow} $} {\bf SU}  \COMMENT{\textit{index fetching}}
 \STATE $l^{\scriptsize \mbox{min}}_m \leftarrow \arg \min_l | r_m[i] - {\boldsymbol{g}_m}  \boldsymbol{s}'_l[i] |^2$
 \STATE $\tilde{{r}}_k[i] = {r}_k[i] - \boldsymbol{h}_k \tilde{\boldsymbol{u}}_k^{\tiny \mbox{ML}}[i]$ \COMMENT{\textit{interference cancellation}}
 \FOR{$lo \leftarrow TI$ \COMMENT{\textit{turbo iterations}}}
 \STATE $\Lambda_1^p[b_{j,k}[i]] \leftarrow $ interleaving aprior, MAP detection
 \STATE $\Lambda_2^p[b_{j,k}[i]] \leftarrow $ deinterleaving aprior, {max-log-MAP decoding}
 \ENDFOR

\ENDFOR
   \STATE Decision of systematic bit is obtained via $\mbox{sign} \{\Lambda_2^p[b_{j,k}[i]]\}$
\end{algorithmic}
\end{algorithm}

\end{document}